\documentclass[a4paper, amsfonts, amssymb, amsmath, reprint, showkeys, nofootinbib, twoside, floatfix,superscriptaddress]{revtex4-1}
\usepackage[english]{babel}
\usepackage{braket}

\usepackage{amsthm}
\usepackage{mathtools}
\usepackage{physics}
\usepackage{xcolor}
\usepackage{graphicx}
\usepackage[left=23mm,right=13mm,top=35mm,columnsep=15pt]{geometry} 
\usepackage{adjustbox}
\usepackage{placeins}
\usepackage[T1]{fontenc}
\usepackage{lipsum}
\usepackage{csquotes}
\usepackage{amsmath,amssymb}
\usepackage{graphicx}
\usepackage{tabularx,booktabs}
\usepackage{tikz}
\usepackage{float}
\usetikzlibrary{quantikz}
\usepackage[utf8]{inputenc}
\usepackage[colorinlistoftodos, color=green!40, prependcaption]{todonotes}
\usepackage[pdftex, pdftitle={Article}, pdfauthor={Author}]{hyperref}
\usepackage[ruled,lined]{algorithm2e}
\usepackage{algpseudocode}
\usepackage{bm}
\usepackage{hyperref}

\usepackage{caption}

\theoremstyle{definition}

\DeclareMathOperator*{\argmin}{arg\,min}
\DeclareMathOperator*{\argmax}{arg\,max}
\usepackage{dsfont}

\usetikzlibrary{shapes.geometric, arrows}
\usetikzlibrary{automata,positioning}

\tikzstyle{startstop} = [rectangle, rounded corners, minimum width=3cm, minimum height=1cm, text centered, draw=black, fill=blue!20]
\tikzstyle{io} = [trapezium, trapezium left angle=70, trapezium right angle=110, minimum width=3cm, minimum height=1cm, text centered, draw=black, fill=blue!30]
\tikzstyle{process} = [rectangle, minimum width=3cm, minimum height=1cm, text centered, draw=black, fill=blue!20]
\tikzstyle{decision} = [diamond, minimum width=3cm, minimum height=1cm, text centered, draw=black, fill= green!30]
\tikzstyle{arrow} = [thick, ->, >=stealth]

\begin{document}

\title{Big data applications on small quantum computers}
\author{Boniface Yogendran}
    \affiliation{University of Edinburgh, School of Informatics, EH8 9AB Edinburgh, United Kingdom}
\date{\today}

\author{Daniel Charlton}
    \affiliation{University of Edinburgh, School of Informatics, EH8 9AB Edinburgh, United Kingdom}
\date{\today}

\author{Miriam Beddig}
    \affiliation{University of Edinburgh, School of Informatics, EH8 9AB Edinburgh, United Kingdom}
\date{\today}

\author{Ioannis Kolotouros}
    \email{i.kolotouros@sms.ed.ac.uk}
    \affiliation{University of Edinburgh, School of Informatics, EH8 9AB Edinburgh, United Kingdom}
\date{\today}

\author{Petros Wallden}
    \email{petros.wallden@ed.ac.uk}
    \affiliation{University of Edinburgh, School of Informatics, EH8 9AB Edinburgh, United Kingdom}
\date{\today}

\begin{abstract}

Current quantum hardware prohibits any direct use of large classical datasets. Coresets allow for a succinct description of these large datasets and their solution in a computational task is competitive with the solution on the original dataset. The method of combining coresets with small quantum computers to solve a given task that requires a large number of data points was first introduced by Harrow \cite{harrow2020small}. In this paper, we apply the coreset method in three different well-studied classical machine learning problems, namely \textit{Divisive Clustering, 3-means Clustering, and Gaussian Mixture Model Clustering}. We provide a Hamiltonian formulation of the aforementioned problems for which the number of qubits scales linearly with the size of the coreset. Then, we evaluate how the variational quantum eigensolver (VQE) performs on these problems and demonstrate the practical efficiency of coresets when used along with a small quantum computer. We perform noiseless simulations on instances of sizes up to 25 qubits on CUDA Quantum and show that our approach provides comparable performance to classical solvers.
\end{abstract}

\maketitle

\section{Introduction}
\label{sec:introduction}

We are currently in the era where computers of $\approx$ 1000 qubits are available and indications of useful quantum computers \cite{bluvstein2023logical} are starting to appear. Although increased in scale, these quantum devices still inherit imperfect operations and short coherence times making them unsuitable for certain quantum algorithms. To address these issues and to check whether these devices can have any valuable advantage over their classical counterparts, people developed hybrid quantum/classical algorithms \cite{cerezo2021variational, bharti2022noisy} that exploit both the computational power of these quantum devices and at the same time the speed and reliability that the classical computers have to offer.

In this framework, the mathematical problem at hand is transformed into an interacting qubit Hamiltonian whose ground state has a one-to-one correspondence with the solution to the problem of interest. As a first step, the quantum computer prepares and measures a (hard to classically simulate) parameterized quantum state. The classical computer then post-processes these measurements and communicates with the quantum device in a continuous feedback loop. This loop usually corresponds to an energy minimization \cite{kandala2017hardware, farhi2014quantum} where the classical computer implements a classical optimization algorithm and the loop terminates when convergence to a minimum occurs. However, further frameworks have been explored such as adiabatically-inspired \cite{kolotouros2023adiabatic, harwood2022improving, keever2023towards} or optimal-transfer inspired \cite{banks2023rapid} algorithms.

Despite their vast number of use cases in quantum chemistry \cite{mcardle2020quantum, barkoutsos2018quantum}, classical optimization \cite{abbas2023quantum,moll2018quantum}, and quantum machine learning \cite{cerezo2022challenges}, these algorithms cannot be directly applied in classical machine learning tasks that require big data sets. Practical machine learning applications require loading enormous data sets of millions (or even billions) of data points that are out of scope for current quantum processors. In \cite{harrow2020small}, Harrow proposed a method for manipulating big data sets on small quantum computers using \emph{coresets}. This technique allows a large collection of data $X$ to be replaced (to within an error $\epsilon$) by a weighted data set $(X',w)$ with a significantly reduced size.

When the task is to minimize the empirical loss over the original data set, the total complexity of the algorithms that are used is significantly reduced if coresets are used \cite{harrow2020small}. An alternative approach is to use a quantum RAM \cite{giovannetti2008quantum}, which allows calls of superposition of data points, and thus algorithms such as Grover \cite{grover1996fast} and HHL \cite{Harrow_2009} can be used. The qRAM has been used for the development of machine learning algorithms such as $q$-means \cite{kerenidis2019q, doriguello2023know} but its development is out of scope for current quantum hardware. 

In the classical machine learning literature, coresets have been extensively studied and recent results have shown that they can be used for efficient training of machine learning models \cite{mirzasoleiman2020coresets} or for improving the performance in training of noisy data \cite{mirzasoleiman2020coresetsn}. In the quantum setting, \cite{xue2023nearoptimal} recently provided a near-optimal quantum algorithm for coreset construction when the task is $k$-means clustering of a dataset. On top of that, prior to our work, \cite{tomesh2021coreset} used coresets to perform 2-means clustering on NISQ computers. In our work, we seek to contribute to the NISQ quantum computing literature, by showing the practicality of coresets in three different machine learning problems.

\textit{Our contributions:}
\begin{itemize}

    \item We investigate how small quantum computers can tackle three different machine learning problems, namely \emph{divisive clustering, 3-means clustering,} and \emph{Gaussian mixture model clustering}.

    \item We provide Hamiltonian formulations for each of the aforementioned problems for which the number of qubits scales linearly with the size of the coreset.

    \item We test the performance of our approach when a small quantum computer works in parallel with a classical computer in a VQE setting compared to classical solvers.

    \item We perform exact noiseless simulations on instances up to 25 qubits on CUDA Quantum.
    
\end{itemize}

\textit{Structure:} In Section \ref{sec:preliminaries} we provide the essential background on Coresets, Variational Quantum Algorithms, and 2-Means Clustering. In Section \ref{sec:problems_formulation} we introduce the three machine-learning problems that we analyze in our manuscript and explain how these can be mapped to ground states of interacting-qubit Hamiltonians. In Section \ref{sec:results} we compare how near-term quantum devices compare with state-of-the-art classical solvers for up to 25 qubits. Finally, in Section \ref{sec:discussion} we conclude with a discussion and an overview of our results, limitations as well as ideas for future work.

\section{Preliminaries}
\label{sec:preliminaries}

\subsection{Datasets and Coresets}\label{dataset_coreset}

Suppose we are given a data set $X = (\boldsymbol{x_1}, \boldsymbol{x_2}, \ldots, \boldsymbol{x_N})$ with $\boldsymbol{x_i} \in \mathbb{R}^d$. The main idea behind coresets is to construct (in an efficient manner) a weighted data set $(X', w)$ of significantly reduced cardinality in order to replace the original data set $X$. This smaller data set can then be used to solve (approximately) a given task related to the original data set $X$.

Consider a finite set of statistical models $Y$. The goal is to find the model $\boldsymbol{y} \in Y$ that describes the given dataset $X$. Any given task (e.g. clustering) can be solved by minimizing a (task-dependent) cost function $\textsc{cost}_X: Y \rightarrow \mathbb{R}_{\geq 0}$. We assume that the cost function can be written as a sum of non-negative functions $g$ over the data set $X$, i.e. 
\begin{equation}
    \textsc{cost}_X(\boldsymbol{y}) = \sum_{\boldsymbol{x}\in X} g(\boldsymbol{x},\boldsymbol{y}), \; g\geq 0.
\end{equation}

\noindent \textbf{Definition 1. ($\epsilon$-coreset \cite{harrow2020small})} Given a data set $X$, an $\epsilon$-\emph{coreset} is a pair $(X',w)$ with $X'\subseteq X$ and $w: X \rightarrow \mathbb{R}_{\geq 0}$ a weight function such that for all $\boldsymbol{y}\in Y$ and $\epsilon > 0$:
\begin{equation}
    |\textsc{cost}_X(\boldsymbol{y}) - \textsc{cost}_{(X',w)}(\boldsymbol{y})| \leq \epsilon |\textsc{cost}_X(\boldsymbol{y})|
\end{equation}
In our experiments, we consider coresets that are constructed entirely in a classical computer and are then used to solve the problem by minimizing the cost function in the quantum computer. In order to construct the coresets, we employed Algorithm 2 of \cite{coreset_ml}. Specifically for $k$-means clustering of a dataset $X$ with cardinality $|X| = d$ and target error $\epsilon$ we require a coreset of size $m = O(\epsilon^{-2}k\log k \min(\frac{k}{\epsilon}, d))$. In our case, however, we first pick the size of the coreset and then assess the performance of our NISQ algorithm with that choice of coreset size. The interested reader can find details on how to construct a coreset in a classical computer in Appendix \ref{appendix:coreset}.

\subsection{Variational Quantum Algorithms}

\begin{figure*}
    \includegraphics[scale=0.5]{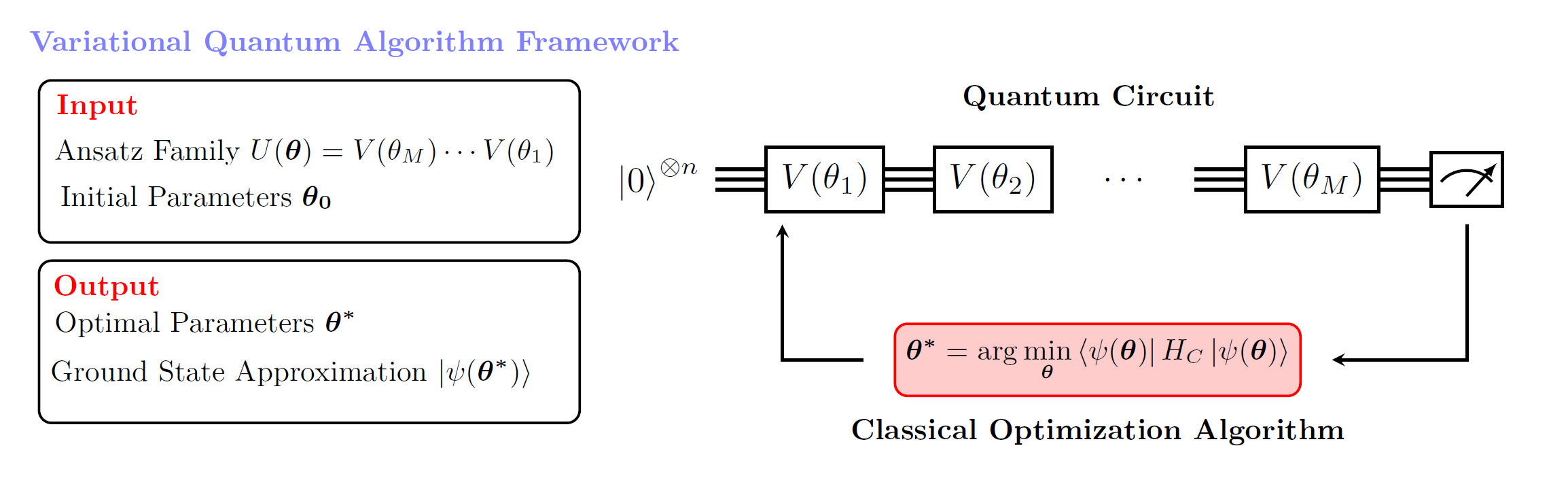}
\caption{General framework of a variational quantum algorithm. The quantum computer iteratively prepares and measures quantum states, and the classical computer employs a classical optimization algorithm to update the parameters (following the direction that minimizes the loss). When the optimization terminates, the algorithm returns a ground state approximation.}
\label{fig:vqas_framework}
\end{figure*}

Variational Quantum Algorithms (VQAs) are a class of hybrid quantum/classical algorithms that are suited for noisy near-term quantum devices. This class of algorithms can tackle any problem (either quantum or classical) that can be cast as an interacting-qubit Hamiltonian. However, in the majority of circumstances they are considered  \textit{heuristics} as they lack strong theoretical guarantees about their performance. You can visualize the general VQA framework in Figure (see Figure \ref{fig:vqas_framework}).

Initially, the given mathematical task is mapped onto an interacting-qubit Hamiltonian decomposed into a linear sum of $L = O(\textrm{poly}(n))$ Pauli strings (i.e. products of Pauli operators on different qubits):
\begin{equation}
    H = \sum_{l=1}^L c_l P_l
\end{equation}
with $c_l \in \mathbb{R}$ and $\norm{P_l} = 1$. Then, the user chooses a parameterized family of gates $U(\boldsymbol{\theta})$ comprised of both single and two-qubit gates. The quantum computer then prepares and measures (a sufficiently large number of times) a parameterized quantum state $\ket{\psi(\boldsymbol{\theta})} = U(\boldsymbol{\theta})\ket{0}$. The measurement outcomes are then post-processed by a classical computer and the parameters are iteratively updated towards the direction that minimizes a loss function, whose minimum will correspond to the solution of the problem of interest. 

Once the classical optimization algorithm has converged, an approximation to the solution is returned and the algorithm terminates. The two main variational quantum algorithms used extensively in the VQA literature are the Quantum Approximate Optimization Algorithm  (QAOA) \cite{farhi2014quantum} and the Variational Quantum Eigensolver (VQE) \cite{kandala2017hardware}. Their main difference is that the QAOA employs a parameterized quantum state that is problem-dependent, compared to VQE where the ansatz family is problem-agnostic. In this manuscript, we employed the latter.

\subsection{2-Means Clustering}
\label{subsec:2-means_clustering}

2-Means clustering aims to group a collection of data $X = (\bm{x_1}, \bm{x_2}, \ldots, \bm{x_N})$ with $\bm{x_i} \in \mathbb{R}^d$ into two separate clusters $S_0, S_1$ so that $S_0 \cup S_1 = X$ and $S_0 \cap S_1 = \varnothing$. Each cluster $S_i$ is described by its cluster center (also called \textit{centroid}) $\bm{\mu}_i = \sum_{j\in S_i}\frac{\bm{x_j}}{|S_i|}$. We then say that a data point $\bm{x_j}$ belongs in cluster $S_i$ if its distance from the centroid $\bm{\mu}_i$ is minimum. The goal of 2-means clustering can then be cast as minimizing the cost function:
\begin{equation}
    C_{\text{2-Means}} = \sum_{i\in S_0}\norm{\bm{x_i} - \bm{\mu_0}}^2 -  \sum_{i\in S_1}\norm{\bm{x_i} - \bm{\mu_1}}^2
\label{eq:2_means_cost_function}
\end{equation}

Finding the optimal solution to the 2-Means clustering problem is $\mathsf{NP}$-hard and so in practice, heuristics such as Lloyd's algorithm \cite{lloyd1982least} are used. These heuristics are known to perform well in practice but are not always guaranteed to find the optimal solution. 

In \cite{tomesh2021coreset}, Tomesh \textit{et al} demonstrated how to perform 2-means clustering using a hybrid quantum/classical architecture. Specifically, they initially showed that one can construct a coreset $(X', w)$ using classical resources and then demonstrated how the 2-means cost function of Eq. \eqref{eq:2_means_cost_function} can be mapped into a MaxCut Hamiltonian corresponding to an all-to-all connected weighted graph. Thus, the Hamiltonian formulation of the 2-means clustering problem is:
\begin{equation}\label{eq:hamiltonian-2means}
    H_{\text{2-Means}} = \frac{1}{2}\sum_{i<j} w_i w_j \bm{x_i}\cdot \bm{x_j} (Z_i Z_j - \mathds{1})
\end{equation}
where $Z_i \equiv \text{diag}(1,-1)$ is the Pauli-$Z$ operator.

\section{Problems Formulation}
\label{sec:problems_formulation}
In this section, we describe how a small quantum computer can tackle the three main big-data applications that are analyzed in our paper. As we already mentioned, these are \emph{divisive clustering}, \emph{3-means clustering}, and \emph{Gaussian mixture models clustering}. 
\subsection{Divisive Clustering}

\subsubsection{Problem Background}\label{divisive-background}

\textit{Hierarchical clustering} refers to a group of unsupervised ML algorithms that can explain both hierarchical and cluster patterns for unlabeled data points. It is further divided into Agglomerative Clustering and Divisive Clustering \cite{kdd_handbook}.

Divisive clustering utilizes a \textit{top-down} approach where all data points are initialized in a single cluster. Then, the algorithm recursively splits the data set into 2 clusters until it reaches singleton clusters \cite{KaufmanR90}. On the other hand, agglomerative clustering is a \textit{bottom-up} hierarchical clustering method that merges the closest clusters recursively until all data points are in a single cluster. It differs from divisive clustering, which recursively splits the entire dataset into smaller clusters, by progressively combining smaller clusters into larger ones based on the proximity between data points \cite{jain1988algorithms}. 

Hierarchical clustering analysis has a vast number of applications and is mostly employed throughout the medical field; it is utilized in big data clinical research, gene expression research, clinical trials, etc  \cite{eisen1998hierarchical, perou2000identification, johnson2012hierarchical}. Here, we focus on how to perform divisive clustering on a NISQ device and show that this is feasible by performing $2$-means clustering iteratively. We choose not to analyze agglomerative clustering, as this requires performing $k$-means clustering for $k>3$ which is beyond the scope of this paper.

\subsubsection{Divisive clustering algorithm}

As explained in Sec. \ref{divisive-background}, in order to perform divisive clustering all data points must be initialized in a single cluster. Then, we recursively perform 2-means clustering until a singleton cluster is reached for all data points. However, in our case, instead of the full dataset, the quantum processor utilizes a coreset that is constructed on the classical computer. Details on how to perform 2-means clustering can be found in Sec. \ref{subsec:2-means_clustering}. The interested reader can find details on how to construct a coreset in Appendix \ref{appendix:coreset} and the pseudocode for divisive clustering in Appendix \ref{appendix:pseudo_divisive}.

The algorithm operates by receiving a coreset as input and generates a data hierarchy as output. Initially, all coreset points are aggregated into a single cluster. Then, at each step, the algorithm assesses whether the number of singleton clusters is equal to the number of coreset points. If this condition is not satisfied, the algorithm proceeds to determine whether the given data points qualify as a singleton cluster. This is determined by counting data points associated with the data provided. If the data point is indeed a singleton cluster, the counter for singleton clusters is updated accordingly. Conversely, if the data point does not constitute a singleton cluster, the coreset undergoes a two-means clustering procedure. This clustering process yields the assignment of data indices to respective groups, based on which the indices are partitioned and stored. The steps above are iteratively repeated on the stored indices. The algorithm terminates once the count of singleton clusters matches the number of coreset points, signaling the completion of the process.

Our approach closely resembles a classical system, with the key distinction lying in the calculation of the cost function. In classical settings, clustering algorithms typically employ a specific cost function. However, our approach introduces a novel methodology for determining the cost function. Notably, previous research by Tomesh et al. \cite{tomesh2021coreset} has demonstrated the equivalence between the classical 2-means cost function and the Hamiltonian of Eq. \eqref{eq:hamiltonian-2means}. Leveraging this established equivalence, we expect our approach to yield similar outcomes to those obtained in classical systems.

\begin{figure*}
\begin{tikzpicture}
\node (img1)  {\includegraphics[scale=0.65]{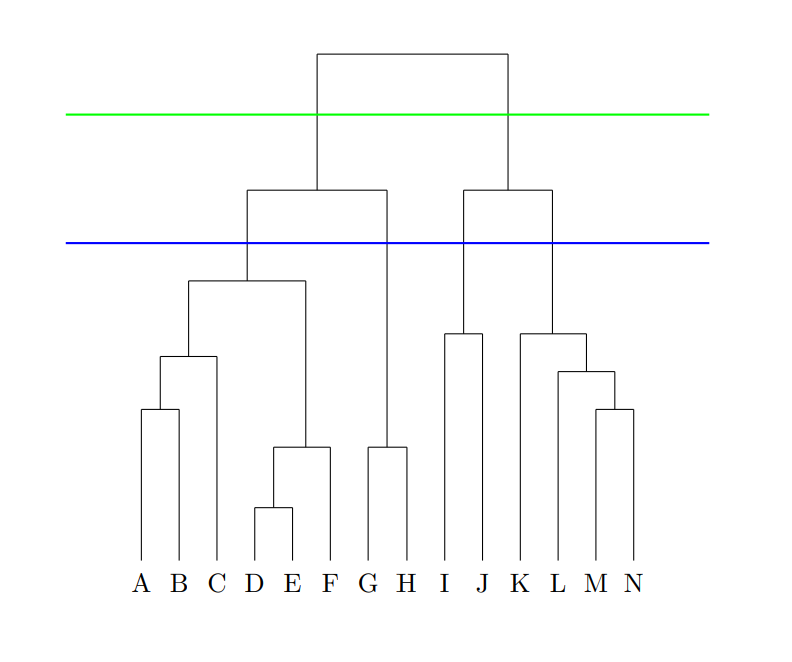}};

\node[right=of img1, yshift=2.5cm, xshift=-1cm] (img2)  {\includegraphics[scale=0.35]{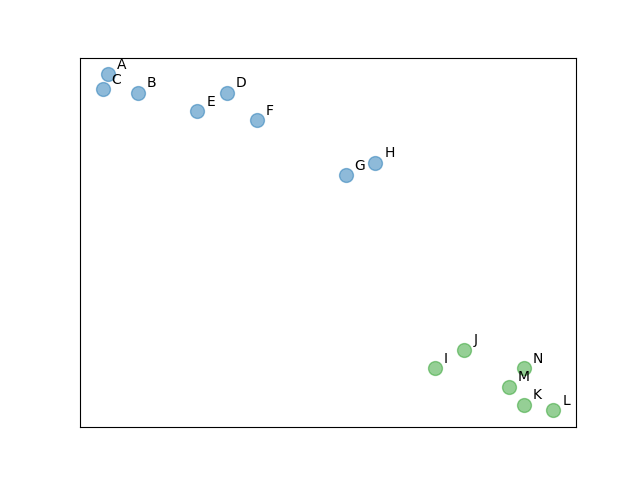}};
\draw[arrow] (3,2.5) -- (4, 3.0);

\node[below=of img2, yshift=1.7cm] (img3)  {\includegraphics[scale=0.35]{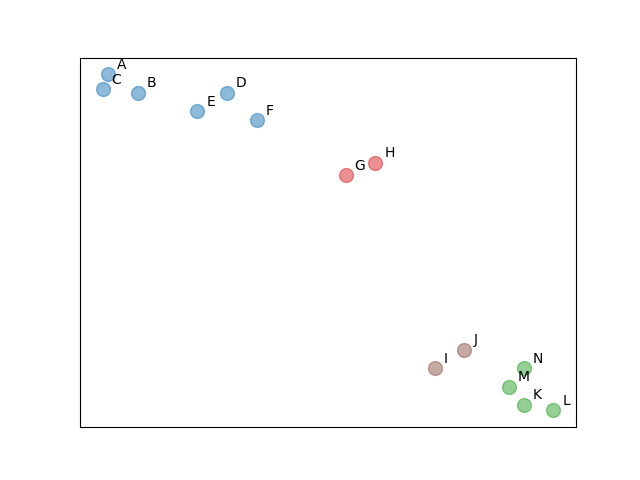}};
\draw[arrow] (3,0.5) -- (4,-0.5);

\end{tikzpicture}
\caption{Hierarchical clustering analysis of synthetic data points presented as a dendrogram (left). By drawing perpendicular lines across the dendrogram, the hierarchical relationships are transformed into data clusters. The number of clusters is determined by the intersections with dendrogram branches. The green and blue horizontal lines (right image) create 2 and 4 clusters, respectively, as indicated. Visual confirmation of data points with the same colors grouped together validates the clustering outcome.}
\label{fig:dendogram_and_scatter}
\end{figure*}

The output of the process is visually represented as a dendrogram in Figure \ref{fig:dendogram_and_scatter}, illustrating the hierarchical arrangement of data points or clusters. By visually examining the data points on the graph and comparing them to the dendrogram, we can observe that data points with shorter branch lengths indicate stronger similarities. To validate the unsupervised learning outcome, we draw a horizontal line intersecting with four branches, indicating the grouping of data points belonging to specific clusters, which are differentiated by color. The scatter plot in Figure \ref{fig:dendogram_and_scatter} further confirms the acceptable outcome, as data points of the same color tend to be located in the same region.

\subsubsection{Cost calculation}
The overall cost for divisive clustering is calculated using the following equation:

\begin{equation}
\label{eq:total_cost_div}
  C_{\text{DC}} = \sum_{l = 1}^T \sum_{i \in S_j} \| x_i - \mu_j \|^2, \;  j \in \{0,1\}  
\end{equation}
where $T$ is the total number of splits, $j$ is the cluster at each split. $x_i$ represents the data point and $\mu_j$ is the centroid of the cluster $j$. $ C_{\text{DC}}$ is the summation of cost at each iteration. It was calculated 10 times and its average is used to analyze the outcome. 

\subsection{3-Means Clustering}

\subsubsection{Problem Background}
The $k$-means clustering problem is a widely applied unsupervised ML technique that is used to group data into $k$ sets (or clusters) by minimizing some distance metric of each point to its nearest cluster center. Finding the cluster centers that minimize the objective function is an NP-hard problem for even the case of $k=2$ \cite{tomesh2021coreset} (see Preliminaries \ref{subsec:2-means_clustering}). Classical algorithms frequently used today, such as LLoyd's algorithm, are heuristics and not guaranteed to find the optimal solution, but perform well in practice \cite{kanungo2002efficient}. Therefore $k$-means satisfies the criterion of a good candidate for VQAs as it is computationally expensive classically. Recently, \cite{jaiswal2023quantum} proposed an approximate polylogarithmic (to the number of data points) algorithm to solve the $k$-means clustering problem, but its implementation requires a qRAM which is out-of-scope of current quantum hardware.

Extending the work of Tomesh \textit{et al}, described in the preliminaries, we investigate the case where $k=3$. We show that the 3-means clustering problem can be restated as a weighted MaxCut instance on a complete graph where the vertices are partitioned into three distinct sets. Following this, we derive a Hamiltonian encoding and present the results of VQE simulations for both random samples and coresets.

\subsubsection{3-Means Clustering Objective Function}

The 3-means clustering problem aims to identify 3 cluster centers that are near the input data. 
As our experiments used weighted coresets, we considered the weighted 3-means clustering problem. Formally the problem is defined as follows: given a dataset $\bm{x}_1,\ldots,\bm{x}_n \in \mathbb{R}^d$ with corresponding weights $w_j \in \mathbb{R}_+$, we aim to find the cluster centers $\bm{\mu}_1, \bm{\mu}_2, \bm{\mu}_3$ that minimize the cost function 

\begin{equation}
\label{eq: weighted 3 means cost}
    C = \sum_{i=1}^{3}\sum_{j \in S_i}w_j\norm{\bm{x}_j - \bm{\mu}_i}^2 
\end{equation}

The index $j$ is assigned to the set $S_i$ if the nearest cluster center to $\bm{x}_j$ is $\bm{\mu}_i$. Analogous to the 2-means clustering case, the partition of the data uniquely determines the cluster centers. Therefore the goal is to find a partition of $[n]$ into 3 sets that minimizes Eq. (\ref{eq: weighted 3 means cost}), thus the objective function can be written as 

\begin{equation}
\label{eq: rewritten 3-means cost}
    \argmin_{S_1, S_2, S_3} \sum_{i = 1}^3 \sum_{j \in S_i} w_j \norm{\bm{x}_j - \bm{\mu}_i} ^ 2.
\end{equation}

The \textit{scatter} of the coreset; a global property of the coreset that measures the squared distance of each point to the overall mean of the coreset $\bm{\mu}$ \cite{tomesh2021coreset}. Notably, the scatter is independent of the partitioning of the coreset, hence it is constant for a given coreset. For the $k$-means case, the scatter can be expressed as 

\begin{align}
\label{eq: k-means scatter}
    \sum_j w_j \norm{\bm{x}_j - \bm{\mu}}^2 &= \sum_{i = 1} ^ k \sum_{j \in S_i} w_j\norm{\bm{x}_j - \bm{\mu}_i}^2 \nonumber \\ 
    &+ \sum_{i = 1} ^ k \sum_{j \in S_i}w_j \norm{\bm{\mu} - \bm{\mu}_i}^2,
\end{align}

This can be shown by expanding the expression for the scatter using the binomial theorem. Alternatively, the result follows from the Law of Total Variance \cite{kriegel2017black}. It can be seen that the first term on the RHS of Eq. \eqref{eq: k-means scatter} is exactly the $k$-means cost function we wish to minimize. As the scatter is constant, we can equivalently maximize the second term on the RHS to find the optimal cluster centers. This term is the weighted between cluster sum of squares \cite{zhao2009sum}.

For a given partition of the data, the cluster centers $\bm{\mu}_i,$  $i=1,2,3$ that minimize Eq. (\ref{eq: weighted 3 means cost}) are given by 

\begin{equation}
\label{eq: 3 means weighted cluster centres}
    \bm{\mu}_i = \frac{1}{W_i}\sum_{j \in S_i} w_j \bm{x}_j,
\end{equation}
where $W_i = \sum_{j \in S_i}w_j$. In addition, we define $W \coloneqq \sum_i W_i$ to be the sum of all edge weights. Applying Eq. \eqref{eq: k-means scatter} for the case $k=3$, the objective function for the 3-means clustering problem can be reformulated as a maximization instance

\begin{equation}
\label{eq: maximisation definition}
    \argmax_{S_1, S_2, S_3} \sum_{i=1}^3 \sum_{j \in S_i} w_j \norm{\bm{\mu} - \bm{\mu}_i} ^ 2.
\end{equation}

\subsubsection{Formulation as a Hamiltonian Problem}
In this section, we will describe how we arrive at a Hamiltonian encoding of Eq. \eqref{eq: rewritten 3-means cost}. We do this by separating \eqref{eq: maximisation definition} into terms that depend on the partition of $[n]$, and those that are independent of this. The full algebra is omitted here for conciseness. We emphasize that in our work we assume that the cluster weights are equal. That is $W_1 = W_2 = W_3 = \frac{W}{3}$. We have included the more general case prior to this point for completeness, moreover, it serves as a starting point for any further investigations where we are not restricted to this assumption. Under this assumption, we can show that

\begin{align}
\label{eq: 3-means clustering final}
    \sum_{i=1}^3 &\sum_{j \in S_i} w_j \norm{\bm{\mu} - \bm{\mu}_i} ^ 2 = 
    \frac{2}{W} \left[ \sum_i w_i^2 \norm{\bm{x}_i}^ 2  \right. \nonumber \\
    &+ 2 \sum_{i < j} w_i w_j \bm{x}_i \cdot \bm{x}_j 
    - 3\left( \sum_{i \in S_1, j \in S_2} w_i w_j \bm{x}_i \cdot \bm{x}_j
      \right. \nonumber \\
      &+ \sum_{i \in S_1, j \in S_3} w_i w_j \bm{x}_i \cdot \bm{x}_j  +   \left. \left. \sum_{i \in S_2, j \in S_3} w_i w_j \bm{x}_i \cdot \bm{x}_j \right) \right].
\end{align}
Here, only the final three terms are dependent on the partition of the coreset. Therefore, after re-scaling, our maximization instance reduces to 

\begin{equation}
    \begin{gathered}
    \argmax_{S_i} - \left(  \sum_{i \in S_1, j \in S_2} w_i w_j \bm{x}_i \right. \cdot \bm{x}_j \\ + \sum_{i \in S_1, j \in S_3} w_i w_j \bm{x}_i \cdot \bm{x}_j  + \left. \sum_{i \in S_2, j \in S_3} w_i w_j \bm{x}_i \cdot \bm{x}_j  \right) .
\end{gathered}
\end{equation}

This is an equivalent formulation of the MaxCut problem on a complete graph, where the vertices are partitioned into three distinct sets. Therefore it is maximized for the weighted MaxCut assignment of the vertices into three sets with edge weights $-w_iw_j \bm{x}_i \cdot \bm{x}_j$. 

To encode the problem as a Hamiltonian we need to introduce a qubit representation of the vertices. Because the coreset is partitioned into three distinct sets it is slightly more difficult to use a binary representation for the graph than it is in the 2-means case. To get around this we introduce a labeling convention, whereby each vertex $\bm{x}_i$ is assigned two labels $y_i, z_i \in \{0,1\}$. The labels indicate the set that each vertex is assigned by the following convention

\begin{gather}
    y_i = 0, \ z_i = 0 \Rightarrow \bm{x}_i \in S_1 \\
    y_i = 1, \ z_i = 0 \Rightarrow \bm{x}_i \in S_2 \\
    z_i = 1 \Rightarrow \bm{x}_i \in S_3.
\end{gather}

The labels allow us to define the function $f(\bm{x}_i, \bm{x}_j)$ which sums over the edges of $G$. If an edge $(i, j)$ crosses the cut then $-w_iw_j \bm{x}_i \cdot \bm{x}_j$ is contributed to the sum, otherwise there is no contribution. We define $f$ as 

\begin{gather}
\label{eq: 3-means Hamiltonian}
         - \sum_{(i, j) \in E(G)} \big [ (1-z_i)  \left((1 - y_i)(y_j + z_j - y_j z_j) \right. \nonumber \\ 
         \left. + y_i(1 - y_j + y_j z_j) \right) 
        + z_i(1 - z_j) \big ] w_i w_j \bm{x}_i \cdot \bm{x}_j
\end{gather}

Recall that the operator $\frac{1}{2}\left(\mathds{1} - Z \right)$ has eigenvalues in the set $\{0, 1\}$, hence we encode each label as Pauli operators as follows

\begin{gather}
    y_i \to \frac{1}{2}\left(\mathds{1} - Z_i \right), \quad z_i \to \frac{1}{2}\left(\mathds{1} - Z_{i^\prime} \right) \\
    y_j \to \frac{1}{2}\left(\mathds{1} - Z_j \right), \quad z_j \to \frac{1}{2}\left(\mathds{1} - Z_{j^\prime} \right).
\end{gather}

The extra subscripts $i^\prime$ and $j^\prime$ are necessary because of the two-qubit encoding of each vertex. The subscripts $i$ and $j$ indicate that the operator acts on the first qubit representing the vertex $v_i$ and $v_j$ respectively, whereas $i^\prime$ and $j^\prime$ subscripts indicate that the operator acts on the second qubit of each respective vertex. Making these substitutions in our expression for $f$ we arrive at the problem Hamiltonian that takes the form

\begin{gather}
    \mathcal{H} = -\frac{1}{8} \sum_{i<j} \left(5 \mathds{1} + Z_{i^\prime} + Z_{j^\prime} -Z_i Z_j - 3Z_{i ^\prime} Z_{j^\prime} \right. \nonumber \\ \left.  - Z_i Z_{i^\prime}Z_j - Z_i Z_j Z_{j^\prime} - Z_i Z_{i^\prime} Z_j Z_{j^\prime}\right) w_i w_j x_i \cdot x_j .
\end{gather}

\subsection{Gaussian Mixture Model Clustering}

\subsubsection{Problem Background}
A Gaussian Mixture Model (GMM) is a statistical model that represents the probability distribution of a dataset as a mixture of multiple Gaussian distributions. The probability density $\varphi$ of such a distribution is 
\begin{align}
\varphi(\boldsymbol{x}) = \sum_{k=1}^K p_k f(\boldsymbol{x}; \boldsymbol{\mu_k}, \Sigma_k), \label{eq:gmm}
\end{align}
where the $p_k$'s are the mixture weights $(p_k \geq 0, \sum_k p_k = 1),$ $K$ is the number of components and $f$ is the probability density of a multivariate normal distribution with mean $\boldsymbol{\mu_k}$ and covariance $\Sigma_k$.


In this work, we focus on GMM clustering. That is, given a dataset $X$ that has been sampled from  Eq. \eqref{eq:gmm}, our goal is to assign to each data point $\boldsymbol{x_i} \in X$ its corresponding normal distribution it is most likely coming from, i.e. we cluster our data into $K$
classes.

For the next paragraphs we introduce the labels $\gamma=(\gamma_1, \dots, \gamma_n)$ where $\gamma_j = i$ if $\boldsymbol{x_j}$  belongs to the $i$-th cluster.
Furthermore we define
\[
S_i := \{ j \in [n] : \gamma_j = i\},
\]
i.e. $S_i$ is the set that contains the indices of all data points that belong to the $i$-th cluster.

\subsubsection{GMM Clustering Objective Function}

We will provide an analysis, on how the aforementioned task can be tackled by a NISQ computer. In our analysis, we will assume that all mixture weights $p_k$ are equal. 
To compute the clustering, we minimize the negative \textit{classification log-likelihood function} 
\cite[§3.6]{gordon1999classification},
\begin{align}
l(\boldsymbol{\theta}, X) &= \text{constant term} + \frac{1}{2}\Bigg( \sum_{j=1}^n \ln(|\Sigma_{\gamma_j}|)  \nonumber\\
&+ (\boldsymbol{x_j} - \boldsymbol{\mu_{\gamma_j}})^\top \Sigma_{\gamma_j}^{-1}(\boldsymbol{x_j} - \boldsymbol{\mu_{\gamma_j}}) \Bigg) .
\label{eq:log_classi_partition}
\end{align}
Here, $\boldsymbol{\theta}$ includes all unknown parameters, i.e. $\boldsymbol{\theta}=(\gamma_1, \dots, \gamma_n, \boldsymbol{\mu_1}, \dots, \boldsymbol{\mu_K}, \Sigma_1, \dots, \Sigma_K)$. 

For any partition of $X$ into $K$ disjoint sets, Eq. \eqref{eq:log_classi_partition} is minimized by substituting the corresponding maximum likelihood estimators for the mean and covariance,
\begin{gather*}
    \boldsymbol{\hat{\mu}_k}=\frac{1}{|S_k|} \sum_{j \in S_k} \boldsymbol{x_j} \\
    \hat{\Sigma}_k=\frac{1}{|S_k|}\sum_{j \in S_k} (\boldsymbol{x_j} - \boldsymbol{\hat{\mu}_k})(\boldsymbol{x_j} - \boldsymbol{\hat{\mu}_k})^\top
\end{gather*}
into Eq. \eqref{eq:log_classi_partition} \cite{gordon1999classification}. Therefore, to find the minimum of  (\ref{eq:log_classi_partition}), it is sufficient to consider all partitions of the dataset $X$ into $K$ disjoint sets. As we previously discussed, finding the minimum over all possible partitions is known to be $\mathsf{NP}$-hard, even for the case $K = 2.$

Since we want to apply the coreset technique, we derive a weighted version 
of Eq. \eqref{eq:log_classi_partition}. Let $(X', w) = (\{\boldsymbol{x_1} , \dots , \boldsymbol{x_m} \}, \{w_1 , \dots , w_m \})$ be a coreset and $(S_1, \dots, S_{K})$ be a clustering of the coreset. Since our goal is to minimize Eq. \eqref{eq:log_classi_partition}, we can drop the constants
and scalars that are independent of the unknown parameters. The cost function can then be defined as
\begin{align} 
    C(\boldsymbol{\theta}) := \sum_{k=1}^{K} \sum_{j \in S_k} w_j
    \Bigg( \ln(|\Sigma_k|) + (\boldsymbol{x_j} - \boldsymbol{\mu_k})^\top \nonumber\\
    \Sigma_k^{-1} (\boldsymbol{x_j} - \boldsymbol{\mu_k})\Bigg)
\label{eq:cost_classi_weighted}
\end{align}
where again $\boldsymbol{\theta} = (S_1, \dots, S_{K}, \boldsymbol{\mu_1}, \dots, \boldsymbol{\mu_K}, \Sigma_1, \dots, \Sigma_K)$ includes the unknown parameters. 
We denote the sum of weights over $S_k$ as \[W_k := \sum_{j \in S_k} w_j\] and the sum of all weights
as \[ W := \sum_{j=1}^m w_j.\]
By setting the partial derivatives of \eqref{eq:cost_classi_weighted} with respect to $\boldsymbol{\mu_k}$ and $\Sigma_k$ to zero, it can be shown that the cost function is minimized if 
\begin{gather*}
    \boldsymbol{\mu_k}=\frac{1}{W_k} \sum_{j \in S_k} w_j\boldsymbol{x_j} \quad\text{and} \\
    \Sigma_k=\frac{1}{W}\sum_{j\in S_k} w_j(\boldsymbol{x_j} - \boldsymbol{\mu_k})(\boldsymbol{x_j} - \boldsymbol{\mu_k})^\top.
\end{gather*}
That is -- as in the unweighted case -- to find the minimum of \eqref{eq:cost_classi_weighted} it is sufficient to consider all possible partitions of the dataset $X$ into $K$ clusters.

\subsubsection{Formulation as a Hamiltonian Problem}

As we already discussed in previous sections, in order to minimize Eq. \eqref{eq:cost_classi_weighted} on a quantum computer, we have to reformulate the cost function as an interacting qubit Hamiltonian, so that its ground state corresponds to the minimum of the original problem. Our encoding
uses ideas from the MaxCut encoding (also found in \cite{lucas2014ising}), adapted to our specific
problem with ideas taken from \cite{tomesh2021coreset} and \cite{scott1971clustering}, \cite{johns1970identifying}. For conciseness a detailed derivation of the encoding is omitted.

For the remaining section, we focus on clustering the dataset into two groups and we assume \emph{the covariances to be equal.} Under this assumption, the cost function becomes
\begin{align}
    C(S_1, S_2) = &\ln(|\Sigma|) \cdot W \nonumber \\
    & + \sum_{k=1}^2 \sum_{j \in S_k}  w_j (\boldsymbol{x_j} - \boldsymbol{\mu_k})^\top \Sigma^{-1} (\boldsymbol{x_j} - \boldsymbol{\mu_k}).
    \label{eq:cost_2_equal_cov}
\end{align}
One can then show that
\begin{alignat}{2}
&\argmin_{S_1, S_2} C(S_1, S_2) \nonumber \\ 
= &\argmax_{S_1, S_2} W_1 W_2 (\boldsymbol{\mu_1} - \boldsymbol{\mu_2})^\top T^{-1} (\boldsymbol{\mu_1} - \boldsymbol{\mu_2}) 
\label{eq:optimisation},
\end{alignat}
where 
$T$ is the total scatter matrix 
\begin{equation*}
    T := \sum_{i=1}^2 \sum_{S_i} w_j (\boldsymbol{x_j} - \boldsymbol{\mu}) (\boldsymbol{x_j} - \boldsymbol{\mu})^\top, 
\end{equation*}
with $\boldsymbol{\mu}$ being the weighted mean of the whole data set, i.e. $\boldsymbol{\mu} = \frac{1}{W} \sum w_i \boldsymbol{x_i}.$

To implement the optimization problem \eqref{eq:optimisation} on a quantum computer, 
we encode a clustering $(S_1, S_2)$ as a computational basis state
$|b \rangle =|b_1...b_m \rangle$, where $b_i = 0$ ($b_i = 1)$ if 
$i \in S_1 \, (i \in S_2)$.
Furthermore, 
we approximate the term 
\begin{align}
W_1 W_2 (\boldsymbol{\mu_1} - \boldsymbol{\mu_2})^\top T^{-1} (\boldsymbol{\mu_1} - \boldsymbol{\mu_2}) \label{eq:equivalent_cost_term}
\end{align}
as a sum of Pauli operators:
We  define the scalar $Z_l$ as $Z_l =+1$ if $l \in S_1$ and $Z_l = -1$ if $l \in S_2$.
The term  \eqref{eq:equivalent_cost_term} can then be approximated via
\begin{align}
W_1 W_2&(\boldsymbol{\mu_1} - \boldsymbol{\mu_2})^\top T^{-1} (\boldsymbol{\mu_1} - \boldsymbol{\mu_2}) \nonumber  \\
&\approx \sum_{i=1}^{m} (1 - \frac{2 Z_i}{W} \sum w_l Z_l )  w_i w_i \boldsymbol{x_i}^\top T^{-1} \boldsymbol{x_i} \nonumber \\
&+ 2 \sum_{i < j} (Z_i Z_j - \frac{Z_i + Z_j}{W} \sum_l w_l Z_l) w_i w_j \boldsymbol{x_i}^\top T^{-1} \boldsymbol{x_j} \nonumber\\
&=: - \mathcal{H}
\label{eq:hamiltonian}
\end{align}

Eq. \eqref{eq:hamiltonian} defines the Hamiltonian $\mathcal{H}$ for our problem where $Z_l$ is replaced by the Pauli-$Z$ operator on the $l$-th qubit.
Note that the number of summands of the Hamiltonian (\ref{eq:hamiltonian}) scales polynomially with the coreset size $m$, and can therefore be efficiently implemented on a quantum device.






% ...


\section{Simulated Experiments}
\label{sec:results}

\begin{figure}
    \centering
        \begin{tikzpicture}
    \node (img1) [scale=0.6]{
         \begin{quantikz}
            \lstick{$\ket{0}$} &\gate{R_y(\theta_0)} &\gate{R_z(\theta_1)} &\ctrl{1} & \qw & \qw &\gate{R_y(\theta_2)} &\gate{R_z(\theta_3)} & \meter{}\\
            \lstick{$\ket{0}$} &\gate{R_y(\theta_4)} &\gate{R_z(\theta_5)} &\gate{x} &\ctrl{1} &\qw &\gate{R_y(\theta_6)} &\gate{R_z(\theta_7)} & \meter{}\\
            \lstick{$\ket{0}$} &\gate{R_y(\theta_8)} &\gate{R_z(\theta_9)} &\qw &\gate{x} &\ctrl{1} &\gate{R_y(\theta_{10})} &\gate{R_z(\theta_{11})} & \meter{} \\
            \lstick{$\ket{0}$} &\gate{R_y(\theta_{12})} &\gate{R_z(\theta_{13})} &\qw &\qw &\gate{x} &\gate{R_y(\theta_{14})} &\gate{R_z(\theta_{15})} & \meter{}
        \end{quantikz} };
    \end{tikzpicture}
    \caption{Parameterized quantum circuit used for the numerical simulations. }
    \label{fig:qcircuit}
\end{figure}
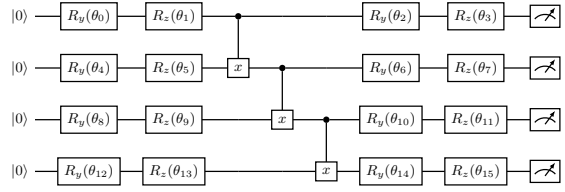

In this section, we provide our numerical results on problem instances consisting of sizes up to 25 qubits. For our simulations, we used VQE with the parameterized architecture seen in Figure \ref{fig:qcircuit}. For the classical optimization part, we used the off-the-shelf gradient-free COBYLA \cite{powell1994direct} optimizer. However, in future work, we intend to use more sophisticated optimization algorithms \cite{stokes2020quantum, kolotouros2023random} that are known to increase the performance of the variational quantum eigensolver (VQE).

For each experiment, we conducted a quantitative analysis by calculating the cost value, i.e. the expectation value of the Hamiltonian (defined on the coreset) of a generated quantum state $\bra{\psi(\boldsymbol{\theta})}\mathcal{H}\ket{\psi(\boldsymbol{\theta})}$. The objective was to compare the output of VQE with three other approaches, serving as proxies to evaluate the performance of our algorithms.

The first approach involves random classification, where the coreset is assigned random labels. This output is anticipated to represent the worst-case scenario. The next algorithm we employed is the Brute Force approach, wherein we systematically explored all possible combinations and selected those with desirable costs. This approach is exhaustive and identifies the true optimum value. The outcome of Brute Force is considered the benchmark for the best possible answer.

Finally, we determined the cost value using Scikit-learn Kmeans implementation \cite{pedregosa2011scikit} which utilizes Lloyd's algorithm. We expect that the VQE results surpass those of the random approach and provide a comparable performance to the classical algorithm. Our goal is not to outperform the best classical algorithm used in practice in these small sizes, but rather, we aim to provide a competitive alternative to the classical counterpart that beats randomly assigned labels and may provide an advantage in the larger scales where the classical algorithms underperform. Finally, we would like to mention that all experiments were repeated 5 times and the optimal value for each of them is presented on Table \ref{results_table}.

\begin{table*}
\begin{center}
\begin{tabular}{||c||c|c|c|c||}
    \hline
      & VQE & Random & Classical & Brute Force \\
     \hline
     Divisive Clustering &1627.19 & 2568.84& 1574.03 & 1450.33\\
     \hline
     3-Means Clustering &4824.61 &5384.88 & 4200.60& 2913.72\\
     \hline
     GMM Clustering & -993.1 &-461 & -992.5& -1004.9\\
    \hline
\end{tabular}
\caption{A total summary of our numerical simulations on instance sizes of 25 qubits.}
\label{results_table}
\end{center}
\end{table*}

\subsection{Divisive clustering}

\begin{figure*}[]
\begin{tikzpicture}
\node (img1)  {\includegraphics[scale=0.35]{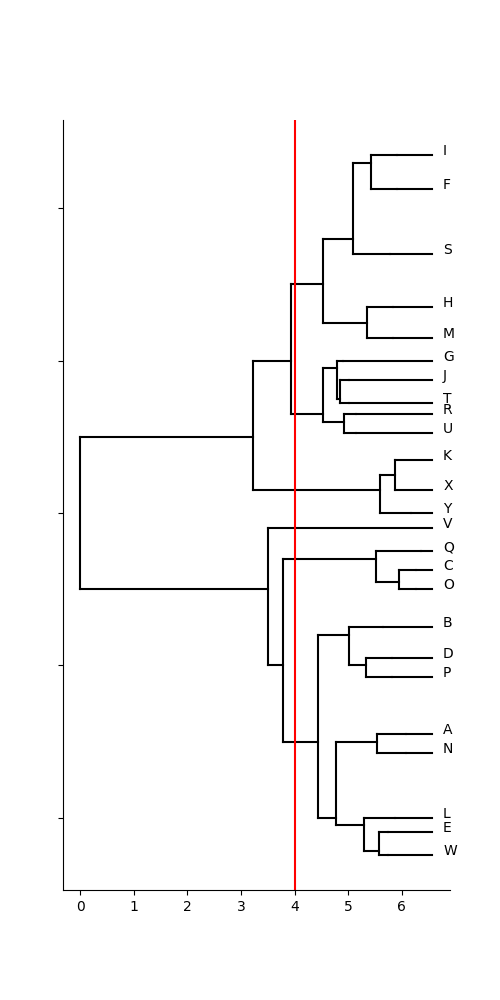}};

\node[right=of img1] (img2)  {\includegraphics[scale=0.55]{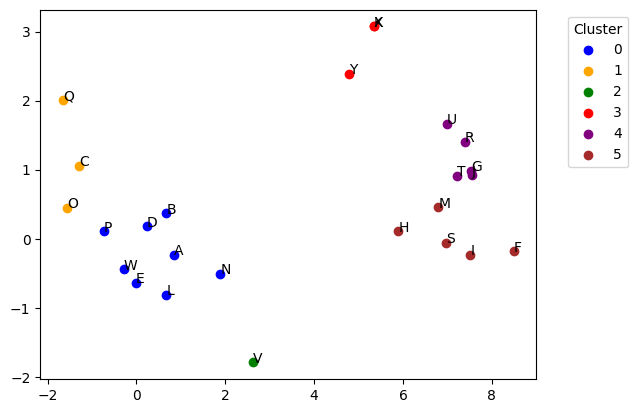}};
\end{tikzpicture}
\caption{On the left side, the dendrogram is illustrated and a vertical line is drawn in order to create 6 clusters. On the right side, we visualize the 6 clusters created by the perpendicular line. The quantum computer is able to correctly cluster the datapoints.}
\label{fig:dc_results}
\end{figure*}

The outcomes of the divisive clustering process are presented in Figure \ref{fig:dc_results} and on the top of Table \ref{results_table}. The outcome corresponds to a 25-qubit instance corresponding to a coreset of equal size. To obtain the cost value of our algorithm for the analysis, we used Eq. \eqref{eq:total_cost_div}. Due to the unsupervised nature of this learning algorithm, the accuracy of the results is challenging to ascertain directly. However, the clustering output serves as a surrogate measure for algorithmic validation. An anticipated outcome involves the observation of logically grouped clusters, discernible by their respective colors. To facilitate cluster creation, a perpendicular line is drawn at a distance of 4, yielding 6 clusters as illustrated in Figure \ref{fig:dc_results}. A visual inspection of this representation confirms the logical grouping of colors within the clusters.

\subsection{3-Means Clustering}
For the 3-means clustering experiment we used a coreset of size 12. The coreset used in this experiment was smaller in size compared to divisive clustering and Gaussian mixture modeling as two qubits are required per coreset point (see Sec \ref{sec:problems_formulation}). We report the costs for both the VQE simulation and classical scikit-learn kmeans implementation on the second line of Table \ref{results_table}. We can see that the cluster centers identified by the VQE simulation produced a cost that was 15\% greater than that of the classical algorithm.



\subsection{GMM Clustering}

Finally for the GMM clustering, we tackled a 25-qubit problem corresponding to a coreset of the same size. For the cost calculation, we used the expectation value of the Hamiltonian in Eq. \eqref{eq:hamiltonian}. The results are presented on the last line of Table \ref{results_table}. As it is clear, our approach (similar to the classical approach) is able to provide a near-optimal solution to the task. 
 


\section{Discussion}
\label{sec:discussion}

In this paper, we addressed how a small (and noisy) quantum computer can tackle three classical machine-learning problems that require large datasets. Specifically, we looked at \emph{divisive clustering}, \emph{3-means clustering} and \emph{gaussian mixture model clustering}. Our starting point has been a classical technique called coresets which was introduced in \cite{harrow2020small}, in which a large dataset $X$ is shrunk down to a weighted dataset $X'$ of significantly reduced cardinality. 

This reduction in the number of data points (which carries information about the large dataset) allows the user to use a small quantum computer and transfer the hardness of the original task on the quantum device. In order to do so, the task on the coreset must be formulated as a ground-state preparation problem. For this reason, we introduced three novel Hamiltonians to solve the aforementioned problems. We show that the number of qubits required to perform divisive clustering and Gaussian mixture modeling in a quantum computer is equal to the size of the coreset. On the other hand, we derive a Hamiltonian that requires a number of qubits that is twice the size of the coreset to perform 3-mean clustering but we also provide alternatives if qutrits are available.

Furthermore, we tested how VQE solvers perform over classical solvers in problems of sizes up to 25 qubits. We performed numerical simulations using CUDA Quantum and observed that the quantum computers can return solutions comparable to their classical counterparts. It is thus interesting to see, how quantum computers perform in instance sizes where classical simulations are prohibitive by testing our proposals on real quantum hardware.

Additionally, the Hamiltonian formulation of the 3-means clustering that we propose, inherits some additional limitations. If the problem graph $G = (V, E)$ has $\abs{V}$ = $n$ we must use a circuit consisting of $2n$ qubits. Quantum resources are precious and ideally, we would like to keep the number of qubits in the circuit to a minimum, furthermore the probability that noise and errors occur in the circuit increases with the number of qubits. We do not need to worry about errors in computation for our simulations, however, we highlight this issue as it becomes a concern when implementing algorithms on quantum hardware. Another issue is the reduction in the size of the coresets that we are able to execute the algorithm on. If we have a given number of qubits at our disposal then the two-label representation effectively forces us to use a coreset of half the size that we could have used if we were conducting 2-means clustering. Generally, the cost of an ML algorithm improves with coreset size, therefore we expect our results to be hindered by this. 

Another observation to consider is that the Hamiltonian contains terms of fourth order. More specifically, there are operators in the Hamiltonian that act on four qubits. This is not an issue for us as such, because we are using a hardware-efficient ansatz, however, it limits the viability of the method when using circuit ansatz families such as the QAOA. For the QAOA the unitary $U(\mathcal{H}, \gamma)$ is dependent on the problem Hamiltonian. Implementing the QAOA to solve this Hamiltonian would therefore require four qubit gates in the circuit. Current quantum hardware has limited connectivity between qubits and for a number of architectures only two-qubit gates can be performed, in many cases this is limited to gates between neighboring qubits. To implement higher-order gates SWAP operations are used to switch the positions of qubits before applying the desired operation. This ultimately leads to increased circuit depth and complexity \cite{wille2019mapping}.

To remedy the issues discussed here, one could consider encoding the vertices of the problem graph using qutrits as this lifts the binary restriction of the qubit labels. We opted not to explore this option in our work for consistency with our other experiments, however, related works have effectively performed simulations using qutrits \cite{bravyi2022hybrid} and have shown very promising results.

At this point, we should stress that even if you are able to find the optimal solution on the coreset, the solution may still differ from the optimal one when considering the full (large) dataset. The reason why this happens is because the reduced dataset carries less information than the full one, with the information increasing with the size of the coreset. There is therefore a trade-off between the accuracy of the solution in the small dataset and the accuracy lost in reducing the large dataset using coresets. Making the coreset of very small size, may result to a problem that a classical computer can brute-force and thus quantum methods offer no advantage. On the other end, if the coreset is of very large size, we would not be able to run a quantum algorithm. Assuming that the quantum algorithm on the small dataset offers some speed-up compared to classical methods for the same task\footnote{Something that can only happen for systems that are beyond the classical simulation limit, i.e. with more than 50 qubits.}, the crucial question to evaluate our approach is if there exists a crossing point where the advantage we get with real devices (that increases with the size of the coreset -- ignoring the effects of noise) outweighs the loss in accuracy due to the reduction of the dataset. Answering this question would involve analyzing the time complexity of the algorithms, taking into account the noise of the system and would highly depend on each application and goes beyond the scope of this work. Our work gives more examples of how this method can be used within machine learning applications and suggests a place where NISQ devices could contribute in giving more accurate solutions. 

\section{Code Availability}
The interested reader can find a Python implementation of our results in \href{https://github.com/Boniface316/VQA}{GitHub} using CUDA Quantum.

\section{Acknowledgements}

PW acknowledges support by EPSRC grants EP/T001062/1, EP/X026167/1 and EP/T026715/1, STFC grant ST/W006537/1 and Edinburgh-Rice Strategic Collaboration Awards.

\bibliographystyle{unsrt}
\bibliography{References}

\onecolumngrid

\appendix

\section{Technical Details}

\subsection{Coreset Construction}
\label{appendix:coreset}

The motivation behind coreset construction was outlined in Section \ref{dataset_coreset}. In order to construct coresets, it is required to define an $(\alpha, \beta)$ bicriterion approximation \cite{BFL16}. Similarly to \cite{coreset_ml}, $D^2$ sampling was used as the initialization step for k-means++ \cite{kmeansplusplus} (see Algorithm \ref{algo-D2-coreset}). For our implementations, $\beta = 2$ was chosen, which corresponds to picking $\beta k = 4$ centroids in the bicriteria approximation.  These are the same values used in \cite{tomesh2021coreset}. 

\begin{center}
        \begin{algorithm}[H]\label{algo-D2-coreset}
        
            \SetAlgoLined

            \SetKwInOut{input}{input}
            \SetKwInOut{output}{output}

            \input{Original data set $X$ \newline
               Iterations $N$ \newline
             }
             
            \output{kmeans++ centroids with the lowest cost \newline}
            
            $X$ $\gets$ normalize each feature of $X$ \;
            best\_cost $\gets \infty$ \;
            \For{i $\gets$ 1,2, $\dots$ $N$}{
                $C \gets D^2$ sampling on $X$ \cite{coreset_ml} \;
                new\_cost $\gets \text{cost}(C, X)$ \;

                \If{new\_cost $<$ best\_cost}{
                   best\_cost $\gets$ new\_cost \;
                    best\_centroids $\gets$ C \;
                }

            }
            
            \Return best\_centroids
        \caption{kmeans++ centroid}
        \end{algorithm}
    \end{center}

After identifying the kmeans++ centroids with the lowest cost, a coreset using Algorithm 2 from \cite{coreset_ml} was created. The algorithm was modified slightly to define the constants and variables to suit the project requirements. $D(\boldsymbol{p}, \boldsymbol{q})$ of Algorithm 2 \cite{BFL16} is treated as the squared Euclidean distance, i.e. for $\boldsymbol{p},\boldsymbol{q}\in \mathbb{R}^n$:
\begin{equation}
    D(\boldsymbol{p},\boldsymbol{q}) = \sum_{i=1}^n (p_i-q_i)^2
\end{equation}

The original data set $X$, kmeans++ centroids $B$ from Algorithm \ref{algo-D2-coreset} and coreset size $m$ are supplied to the algorithm. Initially, the algorithm creates an empty dictionary $\omega$ with a centroid index as the key, and 0 as the value for all keys. Normalizing technique is applied to X. Then for each data point $x \in X$ the Euclidean distance between $x$ and the centroids $b_i\in B$ is calculated. Through this method, the nearest centroid to the data point $x$ is identified. The identified centroid is updated on the dictionary $\omega$. This dictionary will keep count of the number of times a centroid is identified as the nearest centroid in data set X. Then, the sum of the minimum distance between a data point and the centroids is calculated and stored as the variable $sum\_dist$.

The next step of the algorithm is to determine the probability of each data point using the distance. For each data point $x$, it is necessary to calculate the nearest centroid and its distance from it. The probability of each term is calculated by: 

\begin{flalign}\label{eq:prob_eq}
    \text{Pr}(x) = \frac{\left(\underset{b \in B}{min} \; \sqrt{D(x,b)}\right)^2}{2 \sum_{i = 1}^{|X|} \underset{b \in B}{min} \; D(x_i,b)} + \frac{1}{2 |B| \omega \left(\arg \underset{b \in B}{min}  \; \sqrt{D(x,b)}\right) }
\end{flalign}

Where $|B|$ is the total number of centroids in $B$ and $\omega(\arg \min D(x,B))$ is the count for the nearest centroid to $x$. Draw a sample $X'$ of size $m$ from $X$ such that for each $q \in X'$ and $x \in X$ that one has $q = x$ with probability $\text{Pr}(x)$. This sample corresponds to the coreset vectors. The coreset weight is calculated by $w_q = \frac{1}{m \cdot \text{Pr}(q)}$, where $q$ is the data point of the coreset vector $X'$, $m$ is the coreset size, and $\text{Pr}(q)$ is the probability of data point $q$ from the coreset vector $X'$. This is calculated by Eq \ref{eq:prob_eq}.

\begin{center}
    \begin{algorithm}[H]\label{algo-BFL16-coreset}

        \SetAlgoLined

        \SetKwInOut{input}{input}
        \SetKwInOut{output}{output}

        \input{Original data $X$ \newline
               Kmeans++ centroids $B$ \newline
               coreset size $m$
             }

        \output{Coreset vectors $X'$ \newline
                Coreset weights $w$} 
        
        $\omega \gets$ empty dictionary with index of $B$ as key and $0$ as the value \; 
        $X \gets$ normalize each feature of $X$ \;
        $j \gets 0$ \;
        
        \For{$x\in X$}{
            $p_b \gets \arg \underset{b \in B}{min} \; \; \; \sqrt{D(x,b)}$ \;
            $\omega(p_b) \gets \omega(p_b) +1 $ \;
        }
        sum\_dist $\gets \sum_{i = 1}^{|X|} \underset{b \in B}{min} \; \sqrt{D(x,b)}$ \;

        \For {$x \in X$} {
            min\_dist $\gets \underset{b \in B}{min} \; \sqrt{D(x,b)}$ \;
            $p_b \gets \arg \underset{b \in B}{min} \; \; \; \sqrt{D(x,b)}$ \;
            $\text{Pr}(x_i)$ $\gets \frac{min\_dist ^2}{2 \cdot sum\_dist}  + \frac{1}{2|B|\omega(p_b)}$ \;
        }
        
        Pick a sample $X'$ of m points from $X$ such that for each $q \in X'$ and $x \in X$ we have $q = x$ with probability $\text{Pr}(x)$\;

        \For {$x'_i\in X'$} {
            $w[j]  \gets \frac{1}{m \cdot \text{Pr}(x'_i)}$ \;
            $j \gets j + 1$ \;
    
        }

        \Return $X', w$
    \caption{Coreset construction}
    \end{algorithm}
\end{center}

\subsection{Coreset Normalization}

The expectation value of the Hamiltonians derived in Section \ref{sec:problems_formulation} is heavily influenced by coreset vectors and coreset weights. If they are not properly preprocessed, they will negatively impact the outcome of the optimization procedure. We should note here, that there is a distinction between the normalization of the coreset vectors and normalization of the coreset weights.

Initially, the \textit{coreset vectors} are normalized by subtracting each one of its features by their mean and then dividing by the feature's maximum value. This ensures that the Euclidean distance between the two furthest points of each potential cluster is one unit from the origin. Then, the \textit{coreset weights} are normalized by dividing them by the maximum coreset weight value. This will result in a value between 0 and 1. The coreset weight is defined as $\frac{1}{n_c \cdot \text{Pr}(\boldsymbol{x_i})}$, where $n_c$ is the coreset size and $\text{Pr}(\boldsymbol{x_i})$ is the probability of the coreset vector $\boldsymbol{x_i}$. Given that the probability is less than 1, the coreset weight could end up being significantly larger than the edge weight parameters (coreset vectors) $\boldsymbol{x_i} \cdot \boldsymbol{x_j}$. This will significantly influence the cost function and as such normalizing the weights is critical to the performance of the algorithm.

\begin{figure}
    \centering
    \includegraphics[width=0.7\linewidth]{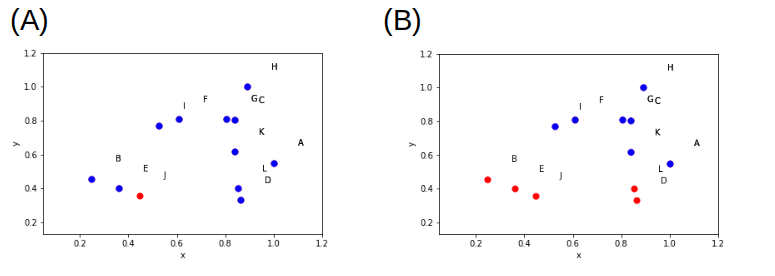}
    \caption{(A) All data points are positive. The clustering outcome is not acceptable as it fails to follow any logic. (B) The outcome of (A) after normalizing the coreset vector. The outcome is acceptable as it follows a logical pattern.}
    \label{fig:all_positive}
\end{figure}

An illustrative example can be visualized in Figure \ref{fig:all_positive} where we perform 2-means clustering on a small dataset. On the left side, we can see the clustering outcome when no normalization is performed while on the right side, we observe the outcome after both coreset-vectors and coreset-weights normalization. It is clear, that the latter results in an accurate result, indicating the importance of preprocessing of the coreset data points.

\section{Divisive Clustering}\label{appendix:pseudo_divisive}

The Divisive Clustering algorithm is presented in Algorithm \ref{algo-DS}. The algorithm takes a coreset $C$, which is a combination of coreset vectors $X'$ and coreset weights $w$ (obtained from Algorithm \ref{algo-BFL16-coreset}), and the coreset size $m$ as inputs. The algorithm initiates various dynamic variables such as the iteration counter $i$, the count of singleton clusters $singleton\_clusters$, an index of data points $index\_values$, and a hierarchical cluster list $hc$. The initial values are set as $i = 0$ and $singleton\_clusters = 0$. The $index\_values$ list stores the indexes of the coreset points used for 2-means clustering, initially containing the indexes of all data points. The $hc$ list is initialized with all data points, as the algorithm starts with all data points forming a single cluster.

The algorithm employs a while loop that runs until the number of singleton clusters is equal to the coreset size ($m$). Within the loop, it extracts the indexes of data points used for 2-means clustering and checks if the number of indexes is 1, indicating a singleton cluster. If so, it increments the count of singleton clusters and proceeds to the next iteration.

If the number of indexes is not 1 (i.e., it's not a singleton cluster), it will extract the coreset vectors and weights that correspond to the indexes. Both coreset weights and the vectors will be normalized. A fully connected graph called $G$ will be created using coreset vectors and weights. MaxCut will be performed on the $G$ and the outcome will stored as a bitstring. Then the bitstring indexes will be used to separate clusters. The indexes corresponding clusters will be appended to the $hc$. Then it will repeat the process until the number of singleton clusters are equal to the size of the coreset. The data hierarchy is presented as a list.

\begin{center}
        \begin{algorithm}[H]\label{algo-DS}
        
            \SetAlgoLined

            \SetKwInOut{Input}{input}
            \SetKwInOut{Parameter}{variables}
            \SetKwInOut{Output}{output}
            
                \Input{Coreset $C$ \newline
                       Coreset size $m$ \newline}

                \Output{Data hierarchy $hc$}

                $i \gets 0$ \hfill \Comment{iteration counter} \;
                $singleton\_clusters \gets 0$ \hfill \Comment{singleton clusters counter} \;
                $index\_values \gets range(m) $ \hfill \Comment{Initialize $index\_values$ with the indexes of all coreset data points}\;
                $hc \gets index\_values$ \hfill \Comment{Create hc as a list containing all data points}\;

                \While{$singleton\_clusters \leq coreset\_size$}
                { 
                    $index\_values \gets hc[i]$ \; 
                    $len\_index\_values \gets $ length($index\_values$) \;
                    \eIf{$len\_index\_values == 1$}{
                        $singleton\_clusters \gets singleton\_clusters + 1$ \;
                        $i \gets i + 1$ \;
                    }
                    {
                    coreset\_vectors $\gets$ coreset vectors of nodes in the $index\_values$ \;
                    coreset\_weights $\gets$ coreset weights of nodes in the $index\_values$ \;
                    coreset\_vectors $\gets$ normalize coreset\_vectors \;
                    coreset\_weights $\gets$ normalize coreset weights \;
                    $G \gets$ Fully connected graph created using coreset\_vectors and coreset\_weights\;
                    
                    Perform MaxCut on the $G$ \;
                    $clusters \gets $ bitstring output from the MaxCut \;
                    
                    \For{$j \gets 0,1$}{
                        $hc.append(index\_values[clusters == j])$ \;
                    }
                    $i \gets i + 1$ \;
                    }}        
            \caption{Divisive clustering}
            \end{algorithm}
\end{center}

\end{document}